# Stronger Reverse Uncertainty Relation for Multiple Incompatible Observables

Xiao Zheng,Ai-Ling Ji,Guo-Feng Zhang[*]

*School of Physics and State Key Laboratory of Software Development Environment,*

*Beihang University, Beijing 100191, China*

**Abstract:** Recently, D. Mondal et.al [Phys. Rev. A. 95, 052117(2017)] creatively introduce a new interesting concept of reverse uncertainty relation which indicates that one cannot only prepare quantum states with joint small uncertainty, but also with joint great uncertainty for incompatible observables. However, the uncertainty upper bound they constructed cannot express the essence of this concept well, i.e., the upper bound will go to infinity in some cases even for incompatible observables. Here, we construct a new reverse uncertainty relation and successfully fix this "infinity" problem. Also, it is found that the reverse uncertainty relation and the normal uncertainty relation are the same in essential, and they both can be unified by the same theoretical framework. Moreover, taking advantage of this unified framework, one can construct a reverse uncertainty relation for multiple observables with any tightness required. Meanwhile, the application of the new uncertainty relation in purity detection is discussed.

**I. Introduction**

Quantum uncertainty relation [1-4], one of the most fundamental differences between quantum and classical mechanics [5-11], indicates that the incompatible observables admit a certain form of uncertainty inequality, which sets the limit on the corresponding measurement precision achievable [12-20]. The initial uncertainty relation deduced by Heisenberg and Robertson, reads [2]:

$$\Delta A^2 \Delta B^2 \geq \left|\tfrac{1}{2i}\langle [A,B]\rangle\right|^2, \qquad (1)$$

where $\Delta A^2 (\Delta B^2)$ stands for the measurement variance of the observable $A(B)$, $[A,B] = AB - BA$, and $\langle [A,B]\rangle = \text{Tr}(\rho[A,B])$ with $\rho$ being the state of the system. Then, Schrödinger [3] constructed a strengthened version of uncertainty relation (1):

$$\Delta A^2 \Delta B^2 \geq \left|\tfrac{1}{2i}\langle [A,B]\rangle\right|^2 + \left|\tfrac{1}{2}\langle \{\check{A},\check{B}\}\rangle\right|^2, \qquad (2)$$

where $\{\check{A},\check{B}\} = \check{A}\check{B} + \check{B}\check{A}$, and $\check{O} = O - \langle O\rangle$.

Both Eq (1) and Eq (2) are uncertainty relations in product form, and these product form uncertainty relations have the trivial problem [21,22]. For example, assuming that the system is in the eigenstates of B and one can easily obtain $\Delta A^2 \Delta B^2 \equiv |\langle [A,B]\rangle/2i|^2 + |\langle \{\check{A},\check{B}\}\rangle/2|^2 \equiv 0$ for finite dimensional Hilbert space even when A and B are incompatible with each other [23-28]. In order to fix this trivial problem, lots of works have been done to investigate the sum form of variance-based uncertainty relations [29-35]. In 2014, Ref. [21] showed that the sum form uncertainty relation they deduced can actually fix the trivial problem of the product form uncertainty relation :

$$\Delta A^2 + \Delta B^2 \geq |\langle\psi|A \pm iB|\psi^\perp\rangle|^2 \pm i\langle[A,B]\rangle, \tag{3}$$

where $|\psi^\perp\rangle$ stands for the state orthogonal to $|\psi\rangle$. However, due to the existence of orthogonal state, the sum uncertainty relation (3) is difficult to apply to high dimension system [29].

Recently, by introducing a new concept of auxiliary operator, Zheng et.al. constructed a unified and exact framework for the variance-based uncertainty relations [22]:

$$\langle \mathcal{F}_1^\dagger \mathcal{F}_1 \rangle \geq \sum_{k=1}^{m} \mathcal{L}_k + \langle \mathcal{F}_{m+1}^\dagger \mathcal{F}_{m+1}\rangle \geq \sum_{k=1}^{m}\mathcal{L}_k \tag{4}$$

where $\mathcal{F}_1 = \sum_{n=1}^{N} x_n \hat{A}_n$, $A_n$ represents the incompatible observables, $N$ is the number of the incompatible observables, $x_n$ stands for random complex number. $\mathcal{L}_k = \left(|\langle[\mathcal{F}_k, \mathcal{O}_k]_\mathcal{G}\rangle|^2 + |\langle\{\mathcal{F}_k, \mathcal{O}_k\}_\mathcal{G}\rangle|^2\right)/4|\langle\mathcal{O}_k^\dagger \mathcal{O}_k\rangle|$ , $[\mathcal{F}_k, \mathcal{O}_k]_\mathcal{G} = \mathcal{F}_k^\dagger \mathcal{O}_k - \mathcal{O}_k^\dagger \mathcal{F}_k$ is the generalized commutator, $\{\mathcal{F}_k, \mathcal{O}_k\}_\mathcal{G} = \mathcal{F}_k^\dagger \mathcal{O}_k + \mathcal{O}_k^\dagger \mathcal{F}_k$ is the generalized anti-commutator [22], $\mathcal{F}_{k+1} = \mathcal{F}_k - \langle\mathcal{O}_k^\dagger \mathcal{F}_k\rangle \mathcal{O}_k / |\langle \mathcal{O}_k^\dagger \mathcal{O}_k\rangle|$ and $\mathcal{O}_k$ represents an arbitrary auxiliary operator. $m$ is the number of the auxiliary operator introduced, and the tightness of the uncertainty relation generally become better and better with the introduction of the auxiliary operators. Moreover, Ref. [22] shows that the uncertainty inequality will become equality when a certain number of auxiliary operators, which satisfy a given condition, are introduced.

This unified framework not only recovers some well-known previous uncertainty relations [22], but also can be used to construct the new uncertainty relations in both product and sum form for two and more incompatible observables with any tightness required. In addition to providing tighter lower bound, this unified framework can also be used to construct the stronger uncertainty relation to fix the trivial problems in both sum form and product form uncertainty relations.

Recently, Ref. [36] showed that, in addition to the uncertainty relation, there exists a new concept of reverse uncertainty relation, which indicates that one cannot only prepare quantum states with joint sharp value, but also with joint great uncertainty for incompatible observables. Therefore, the sum of variances for the incompatible observables has an upper bound at a quantum level. Similar to uncertainty relation, the reverse one is also a unique feature of quantum mechanics. The uncertainty relation indicates the impossibility to make the variance of incompatible observables arbitrarily small at same time, and the reverse uncertainty relation expresses the maximum extent that the variance for the incompatible observables can achieve. The reverse uncertainty relation is firstly deduced by Debasis Mondal in Ref [36], which reads:

$$\Delta A^2 + \Delta B^2 \leq \frac{2\Delta(A-B)^2}{\left[1-\frac{cov(A,B)}{\Delta A.\Delta B}\right]} - 2\Delta A.\Delta B, \tag{5}$$

where $cov(A,B) = \langle\{A,B\}\rangle/2 - \langle A\rangle\langle B\rangle$. However, the reverse uncertainty relation (5) is trivial in some cases, i.e. the upper bound of Eq. (5) approaches infinity when $\Delta A.\Delta B$ exactly is equal to $cov(A,B)$, and it is obvious that the infinite upper bound cannot capture the physical essence of the reverse uncertainty relation.

In this paper, we mainly construct a new reverse uncertainty relation to fix the infinity-upper-bound problem from a new perspective, and show that the new reverse uncertainty relation can also be unified by the uncertainty relation framework constructed in Ref. [22], which indicates that the uncertainty relation and the reverse one are essentially the same. What's more, taking advantage of the unified framework, one can construct a series of reverse uncertainty relation with any tightness required. The outline of the paper is as follows. In Sec. II, we construct a new reverse uncertainty relation and fix the trivial problem in the traditional reverse uncertainty relation. Also, the application of the new uncertainty relation in the purity detection of the quantum state is demonstrated. In Sec. III, we discuss the relationship between the new reverse uncertainty relation and the unified framework, and also use the unified formwork to construct a series of reverse uncertainty relation for N incompatible observables with any tightness one needed. Finally, Sec. IV is devoted to the discussion and conclusion.

## II.  Construction of Variance-Based Reverse Uncertainty Relation

### A.  Mathematical Preparation

Assuming that the state of the system is $\rho$, one can define a bilinear operator function, which reads:

$$F(\mathcal{A},\mathcal{B}) = \text{Tr}(\rho.\mathcal{A}^\dagger.\mathcal{B}), \tag{6}$$

where $\mathcal{A}$ and $\mathcal{B}$ stand for two arbitrary operators. It is easy to prove that the bilinear function $F(\mathcal{A}, \mathcal{B})$ has the following two properties:

$$F(\mathcal{A}, \mathcal{A}) \geq 0. \tag{7a}$$

$$F(\mathcal{A}, \mathcal{A})F(\mathcal{B}, \mathcal{B}) \geq |F(\mathcal{A}, \mathcal{B})|^2. \tag{7b}$$

*Proof*: It should be noted that the operators $\mathcal{A}$ and $\mathcal{B}$ may be non-Hermitian. Firstly, assuming that the quantum state of the system is a pure one, which means the corresponding density operator can be written as $\rho = |\varphi\rangle\langle\varphi|$, and one thus has:

$$F(\mathcal{A}, \mathcal{A}) = \mathrm{Tr}(\rho \cdot \mathcal{A}^\dagger \cdot \mathcal{A}) = \langle\varphi|\mathcal{A}^\dagger \cdot \mathcal{A}|\varphi\rangle. \tag{8}$$

Decompose the operator $\mathcal{A}$ into:

$$\mathcal{A} = \sum_k (a_{kr} + ia_{ki})|a_k\rangle\langle a_k|,$$

where $a_{kr} + ia_{ki}$ is the eigenvalue of the matrix $\mathcal{A}$, $|a_k\rangle$ is the corresponding eigenstate, $a_{kr}$ and $a_{ki}$ represent the real and imaginary parts of the eigenvalue, respectively. Then, one has:

$$\mathcal{A}|\varphi\rangle = \sum_k (a_{kr} + ia_{ki})|a_k\rangle\langle a_k|\varphi\rangle = \sum_k (\overline{a_{kr}} + i\overline{a_{ki}})|a_k\rangle,$$

where $\overline{a_{kr}}$ and $\overline{a_{ki}}$ represent the real and imaginary parts of $\langle a_k|\varphi\rangle(a_{kr} + ia_{ki})$, namely, $\overline{a_{kr}} + i\overline{a_{ki}} = \langle a_k|\varphi\rangle(a_{kr} + ia_{ki})$. By straight calculation, one can obtain:

$$\langle\varphi|\mathcal{A}^\dagger \cdot \mathcal{A}|\varphi\rangle = \sum_k (\overline{a_{kr}}^2 + \overline{a_{ki}}^2) \geq 0.$$

Taking the same method, one can obtain the same conclusion for the quantum mixed state $\rho = \sum_n p_n|\varphi_n\rangle\langle\varphi_n|$ with $p_n$ being the weight of the state $|\varphi_n\rangle$ in the mixed state. Therefore, we have completed the proof of Eq.(7a).

As for Eq.(7b), assume that

$$\mathcal{C} = \mathcal{A} - \mathcal{A}_v = \mathcal{A} - \frac{F(\mathcal{A}, \mathcal{B})}{F(\mathcal{B}, \mathcal{B})}\mathcal{B},$$

where $\mathcal{A}_v$ is defined as $\mathcal{A}_v = F(\mathcal{A}, \mathcal{B})\mathcal{B}/F(\mathcal{B}, \mathcal{B})$. Then, one has:

$$F(\mathcal{C}, \mathcal{B}) = F(\mathcal{A}, \mathcal{B}) - \frac{F(\mathcal{A}, \mathcal{B})}{F(\mathcal{B}, \mathcal{B})}F(\mathcal{B}, \mathcal{B}) = 0.$$

After a simple deformation, one can obtain:

$$F(\mathcal{A}, \mathcal{A}) = F\left(\mathcal{C} + \frac{F(\mathcal{A}, \mathcal{B})}{F(\mathcal{B}, \mathcal{B})}\mathcal{B}, \mathcal{C} + \frac{F(\mathcal{A}, \mathcal{B})}{F(\mathcal{B}, \mathcal{B})}\mathcal{B}\right) = F(\mathcal{C}, \mathcal{C}) + \left|\frac{F(\mathcal{A}, \mathcal{B})}{F(\mathcal{B}, \mathcal{B})}\right|^2 F(\mathcal{B}, \mathcal{B}).$$

$$= F(\mathcal{C}, \mathcal{C}) + \frac{|F(\mathcal{A}, \mathcal{B})|^2}{F(\mathcal{B}, \mathcal{B})} \tag{9}$$

tBased on the proof above and Eq. (7a), Eq. (7b) is obtained.

$$F(\mathcal{A},\mathcal{A})F(\mathcal{B},\mathcal{B}) = F(\mathcal{C},\mathcal{C})F(\mathcal{B},\mathcal{B}) + |F(\mathcal{A},\mathcal{B})|^2 \geq |F(\mathcal{A},\mathcal{B})|^2. \qquad (10)$$

### B. Construction of the Reverse Uncertainty Relation

Assume that the system is in the maximum-mixed state, and Eq.(7b) then turns into:

$$F_{mx}(\mathcal{A},\mathcal{A})F_{mx}(\mathcal{B},\mathcal{B}) \geq |F_{mx}(\mathcal{A},\mathcal{B})|^2 \qquad (11)$$

where $F_{mx}(\mathcal{A},\mathcal{B}) = \text{Tr}(\rho_{mx} \cdot \mathcal{A}^\dagger \cdot \mathcal{B})$, with $\rho_{mx} = I/d$ being the maximum-mixed state and $d$ being the dimension of the system. Taking $\mathcal{A} = \rho$ and $\mathcal{B} = (\check{A} \pm i\check{B})(\check{A} \mp i\check{B})$ into Eq.(11), one can obtain:

$$F_{mx}\left(\rho,(\check{A} \pm i\check{B})(\check{A} \mp i\check{B})\right) \leq \sqrt{F_{mx}(\rho,\rho)}\sqrt{F_{mx}\left((\check{A} \pm i\check{B})(\check{A} \mp i\check{B}),(\check{A} \pm i\check{B})(\check{A} \mp i\check{B})\right)} \qquad (12)$$

where A and B are two arbitrary incompatible observables. Taking advantage of the definition of the bilinear operator function (6), one has:

$$F_{mx}\left(\rho,(\check{A} \pm i\check{B})(\check{A} \mp i\check{B})\right) = \frac{1}{d}\text{Tr}\left(\rho \cdot (\check{A} \pm i\check{B}) \cdot (\check{A} \mp i\check{B})\right) = \frac{1}{d}(\Delta A^2 + \Delta B^2 \mp i\langle[A,B]\rangle) \qquad (13a)$$

$$F_{mx}(\rho,\rho) = \frac{1}{d}\text{Tr}(\rho^2) \qquad (13b)$$

Submitting Eq. (13) and Eq. (12), one can obtain a new reverse uncertainty relation:

$$\Delta A^2 + \Delta B^2 \leq \sqrt{\text{Tr}(\rho^2)}\sqrt{dF_{mx}\left((\check{A} \pm i\check{B})(\check{A} \mp i\check{B}),(\check{A} \pm i\check{B})(\check{A} \mp i\check{B})\right)} \pm i\langle[A,B]\rangle \qquad (14)$$

Taking advantage of Eq.(7a), one has $F\left((\check{A} \mp i\check{B}),(\check{A} \mp i\check{B})\right) = \text{Tr}\left(\rho(\check{A} \pm i\check{B})(\check{A} \mp i\check{B})\right) \geq 0$, which means that the matrix $(\check{A} \pm i\check{B})(\check{A} \mp i\check{B})$ is non-negative. Ulitizing the non-negativity of $(\check{A} \pm i\check{B})(\check{A} \mp i\check{B})$, one thus can obtain the following inequality:

$$\sqrt{dF_{mx}\left((\check{A} \pm i\check{B})(\check{A} \mp i\check{B}),(\check{A} \pm i\check{B})(\check{A} \mp i\check{B})\right)} \leq dF_{mx}\left((\check{A} \pm i\check{B})(\check{A} \mp i\check{B})\right) \qquad (15)$$

Taking Eq. (15) into Eq. (14) one can obtain another new inverse uncertainty relation:

$$\Delta A^2 + \Delta B^2 \leq \sqrt{\text{Tr}(\rho^2)}dF_{mx}\left((\check{A} \pm i\check{B})(\check{A} \mp i\check{B})\right) \pm i\langle[A,B]\rangle. \qquad (16)$$

Taking advantage of $\text{tr}(\rho^2) \leq 1$, one can deduce the third new reverse uncertainty relation:

$$\Delta A^2 + \Delta B^2 \leq dF_{mx}\left((\check{A} \pm i\check{B})(\check{A} \mp i\check{B})\right) \pm i\langle[A,B]\rangle. \qquad (17)$$

In order to compare the new reverse uncertainty relation with the traditional one, we denote the corresponding lower bound as:

$$U_{old} = \frac{2\Delta(A-B)^2}{\left[1-\frac{cov(A,B)}{\Delta A.\Delta B}\right]} - 2\Delta A.\Delta B, \qquad (18)$$

$$U_{new} = dF_{mx}\left((\breve{A} \pm i\breve{B})(\breve{A} \mp i\breve{B})\right) \pm i\langle[A,B]\rangle. \tag{19}$$

The state of the system is chosen as

$$\rho_{ini} = \frac{1}{2}\left(I + \cos\left(\frac{\theta}{2}\right)\sigma_x + \cos\left(\frac{\varphi}{2}\right)\sin\left(\frac{\theta}{2}\right)\sigma_y + \sin\left(\frac{\varphi}{2}\right)\sin\left(\frac{\theta}{2}\right)\sigma_z\right), \tag{20}$$

where $I$ stands for the identity matrix, $\sigma_x, \sigma_y, \sigma_z$ are standard Pauli matrices and $\theta \in (0, 2\pi)$. Taking the incompatible observables as $A = \sigma_x$ and $B = \sigma_z$, one then can obtain:

$$U_{new} = 2\left(2 \pm \cos\left(\frac{\theta}{2}\right)^2 + \sin\left(\frac{\theta}{2}\right)\left(\cos(\varphi) + \sin\left(\frac{\theta}{2}\right)\sin(\varphi)^2\right)\right). \tag{21a}$$

$$\Delta A . \Delta B = \sqrt{\sin\left(\frac{\theta}{2}\right)^2 - \sin\left(\frac{\theta}{2}\right)^4 \sin(\varphi)^2}. \tag{21b}$$

$$\Delta(A - B)^2 = 2 - \cos\left(\frac{\theta}{2}\right)^2 + \sin(\varphi)\left(\sin(\theta) - \sin\left(\frac{\theta}{2}\right)^2 \sin(\varphi)\right). \tag{21c}$$

$$cov(A, B) = -\frac{1}{2}\sin(\theta)\sin(\varphi). \tag{21d}$$

$$\Delta A^2 + \Delta B^2 = \sin\left(\frac{\theta}{2}\right)^2 + \left(1 - \sin\left(\frac{\theta}{2}\right)^2 \sin(\varphi)^2\right) = 1 + \cos(\varphi)^2 \sin\left(\frac{\theta}{2}\right)^2. \tag{21e}$$

Taking Eq.(21) into Eq.(18), one can obtain Fig.1. As shown in Fig.1, the reverse uncertainty relation (5) is trivial when $\theta \approx 7\pi/4$, i.e. the corresponding upper bound $U_{old}$ trends to infinity, and this trivial problem can be well fixed by the new reverse uncertainty relation (17).

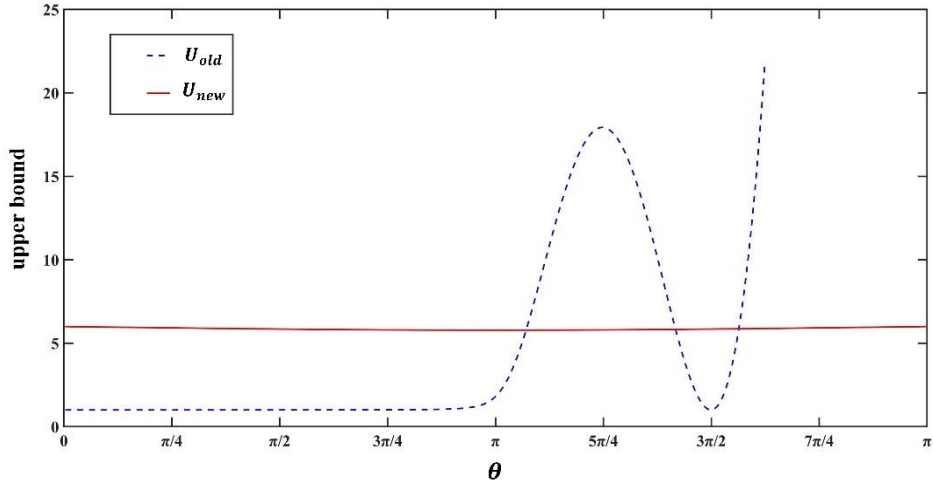

**Fig 1** The evolution of the $U_{old}$ and $U_{new}$ with $\theta$, here $\varphi$ is taken as $(\pi/2 - 1/10)$.

### C. Estimation of purity

The applications of the new obtained uncertainty in the estimation of purity will be discussed in this subsection. For a density matrix $\rho$, the state is a pure when $\text{Tr}(\rho^2) = 1$, and $0 < \text{Tr}(\rho^2) < 1$ for the mixed one. Denoting $\text{Tr}(\rho^2)$ by $Pu$, one can deduce that the bigger the value of $Pu$ is, the greater the purity of $\rho$ is. Therefore, the value of $Pu$ can be employed to quantify the purity of the system [37,38]. According to Ref. [38], the purity is important in quantum information processing and usually difficult to be detected. After a simple deformation of Eq. (16), one can deduce:

$$Pu \geq \left(\frac{(\Delta A^2 + \Delta B^2 \mp i\langle [A,B]\rangle)}{dF_{mx}\left((\breve{A} \pm i\breve{B})(\breve{A} \mp i\breve{B})\right)}\right)^2.$$

Thus, the lower bound of the purity can be easily estimated by detecting the corresponding uncertainty and doing some simple calculations, which is meaningful for an experimenter to estimate the purity of a prepared state in the quantum information processing.

### III. Reverse Uncertainty Relation and the Unified Framework

This section is mainly used to discuss the relationship between uncertainty relation and the reverse one. Take $m = 1$, and the unified uncertainty relation framework (4) then turns into:

$$\langle \mathcal{F}_1^\dagger \mathcal{F}_1 \rangle \geq \frac{|\langle [\mathcal{F}_1, \mathcal{O}_1]_\mathcal{G}\rangle|^2 + |\langle \{\mathcal{F}_1, \mathcal{O}_1\}_\mathcal{G}\rangle|^2}{4|\langle \mathcal{O}_1^\dagger \mathcal{O}_1\rangle|} \tag{22}$$

Assume that the system is in the maximum-mixed state, and the Eq.(22) becomes:

$$\langle \mathcal{F}_1^\dagger \mathcal{F}_1 \rangle_{mx} \geq \frac{|\langle [\mathcal{F}_1, \mathcal{O}_1]_\mathcal{G}\rangle_{mx}|^2 + |\langle \{\mathcal{F}_1, \mathcal{O}_1\}_\mathcal{G}\rangle_{mx}|^2}{4|\langle \mathcal{O}_1^\dagger \mathcal{O}_1\rangle_{mx}|} \tag{23}$$

where $\langle ... \rangle_{mx}$ means the expected value in the maximum-mixed state $\rho_{mx}$. Taking $\mathcal{F}_1 = (\breve{A} \pm i\breve{B})(\breve{A} \mp i\breve{B})$ and the auxiliary operator $\mathcal{O}_1 = \rho$ into Eq.(23), one can obtain:

$$\langle (\breve{A} \pm i\breve{B})(\breve{A} \mp i\breve{B})(\breve{A} \pm i\breve{B})(\breve{A} \mp i\breve{B})\rangle_{mx} \geq \frac{\left|\langle [(\breve{A} \pm i\breve{B})(\breve{A} \mp i\breve{B}), \rho]_\mathcal{G}\rangle_{mx}\right|^2 + \left|\langle \{(\breve{A} \pm i\breve{B})(\breve{A} \mp i\breve{B}), \rho\}_\mathcal{G}\rangle_{mx}\right|^2}{4|\langle \rho\rho\rangle_{mx}|} \tag{24}$$

After some simple calculation, one has:

$$\left|\langle [(\breve{A} \pm i\breve{B})(\breve{A} \mp i\breve{B}), \rho]_\mathcal{G}\rangle_{mx}\right|^2 + \left|\langle \{(\breve{A} \pm i\breve{B})(\breve{A} \mp i\breve{B}), \rho\}_\mathcal{G}\rangle_{mx}\right|^2$$
$$\leq 4\langle (\breve{A} \pm i\breve{B})(\breve{A} \mp i\breve{B})(\breve{A} \pm i\breve{B})(\breve{A} \mp i\breve{B})\rangle_{mx}|\langle \rho\rho\rangle_{mx}| \tag{25}$$

Submitting $\left|\langle [(\breve{A} \pm i\breve{B})(\breve{A} \mp i\breve{B}), \rho]_\mathcal{G}\rangle_{mx}\right|^2 + \left|\langle \{(\breve{A} \pm i\breve{B})(\breve{A} \mp i\breve{B}), \rho\}_\mathcal{G}\rangle_{mx}\right|^2 = 4(\Delta A^2 + \Delta B^2 \mp i\langle [A,B]\rangle)^2/d^2$ into Eq.(25), one then can obtain the reverse uncertainty relation (14). That is to say, similar to normal uncertainty relation, the reverse uncertainty relation can be deduced by introducing auxiliary operator, which means that both the uncertainty relation and the reverse one can be unified by the same theoretical framework, and the two relation both are inherent properties of quantum mechanics.

Also, similar to the case of uncertainty relation, to construct stronger reverse uncertainty relation, one can use the unified framework to introduce m auxiliary operators, with $m \geq 2$:

$\Delta A^2 + \Delta B^2 \leq$

$$\sqrt{\text{Tr}(\rho^2)}\sqrt{dF_{mx}\left((\breve{A}\pm i\breve{B})(\breve{A}\mp i\breve{B}),(\breve{A}\pm i\breve{B})(\breve{A}\mp i\breve{B})\right) - d\sum_{k=2}^{m}\mathcal{L}_k^{mx}} \pm i\langle[A,B]\rangle \quad (26)$$

where $\mathcal{L}_k^{mx}$, the embodiment of the k-th auxiliary operator in the reverse uncertainty relation, is the value of $\mathcal{L}_k$ in the maximum-mixed state $\rho_{mx}$. According to the expression of $\mathcal{L}_k$, one can easily deduce that the item $\sum_{k=2}^{m}\mathcal{L}_k^{mx} \geq 0$. Then, one can obtain that the upper bound become smaller and smaller, which means the tightness of the reverse uncertainty relation become better and better, with the introduction of the auxiliary operators, as shown in Fig.2.

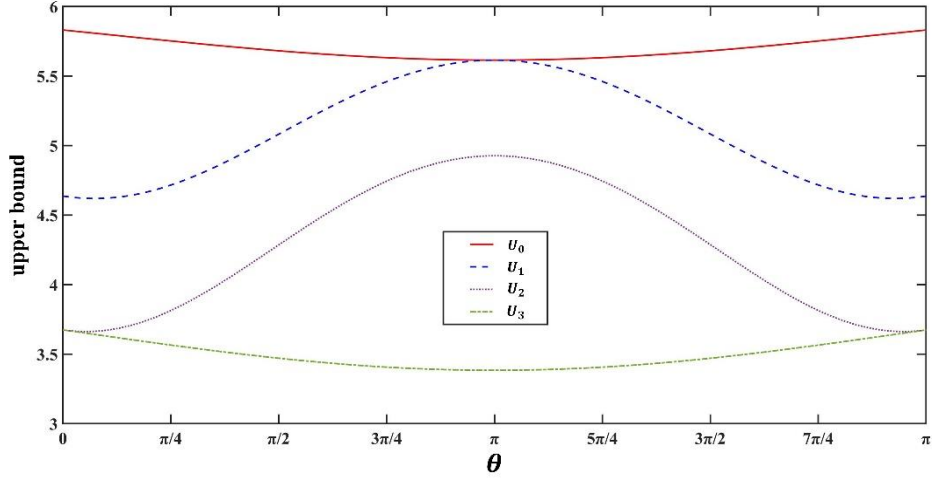

**Fig 2** Evolution of the $U_0$, $U_1$, $U_2$ and $U_3$ with $\theta$. Here the state is taken as $\rho_{ini}$, $\varphi = (\pi/2 - 1/10)$, $U_i$ represents the upper bound with $i + 1$ auxiliary operators ($i = 0,1,2,3$), $\mathcal{O}_1 = \rho$, $\mathcal{O}_2 = \sigma_x$, $\mathcal{O}_3 = \sigma_y$, and $\mathcal{O}_4 = \sigma_z$.

Moreover, we will show in the following that taking the unified framework, we can construct the reverse uncertainty relation for multiple observables. Taking $\mathcal{F}_1 = \left(\sum_{k=1}^{N}e^{-i\theta_k}\breve{A}_k\right)\left(\sum_{l=1}^{N}e^{i\theta_l}\breve{A}_l\right)$ and $\mathcal{O}_1 = \rho$ into the unified framework, one can obtain the reverse uncertainty relation for N incompatible observables with $m$ auxiliary operator:

$\sum_{j=1}^{N}\Delta A_j^2 \leq -\sum_{k<l}\langle\{e^{i\theta_k}\breve{A}_k, e^{i\theta_l}\breve{A}_l\}_\mathcal{G}\rangle +$

$$\sqrt{d\text{Tr}(\rho^2)\left(F_{mx}\left[\left(\sum_{k=1}^{N}e^{-i\theta_k}\breve{A}_k\right)\left(\sum_{l=1}^{N}e^{i\theta_l}\breve{A}_l\right)\left(\sum_{k=1}^{N}e^{-i\theta_k}\breve{A}_k\right)\left(\sum_{l=1}^{N}e^{i\theta_l}\breve{A}_l\right)\right] - \sum_{k=2}^{m}\mathcal{L}_k^{mx}\right)} \quad (27)$$

where $\theta \in [0, 2\pi]$ should be chosen to minimize the upper bound, $\mathcal{L}_k^{mx} = (|\langle[\mathcal{F}_k, \mathcal{O}_k]_\mathcal{G}\rangle_{mx}|^2 + |\langle\{\mathcal{F}_k, \mathcal{O}_k\}_\mathcal{G}\rangle_{mx}|^2)/4|\langle\mathcal{O}_k^\dagger\mathcal{O}_k\rangle_{mx}|$, and $\mathcal{O}_k$ is k-th auxiliary operator, which can be chosen as any operator. Similar to the discussion in Eq. (26), we can deduce that $\sum_{k=2}^{m}\mathcal{L}_k^{mx} \geq 0$. That is to say, one can construct the reverse uncertainty relation with any tightness required by introducing more auxiliary operators.

**IV. CONCLUSIONS**

In conclusion, we have constructed a new reverse uncertainty relation, the upper bound of which is guaranteed to be non-trivial, and the obtained relation thus fix the "infinity" problem of the existing ones. Also，it is found that the reverse uncertainty relation and the normal ones can be unified to the same framework, which indicates that the two types of uncertainty relations are essentially the same. Moreover, we show that, taking advantage of this unified framework, one can construct the multi-observable reverse uncertainty relation with any tightness needed.

**Acknowledgments**

This work is supported by State Key Laboratory of Software Development Environment and National Natural Science Foundation of China (Grant No. 11574022) .